\documentclass[aps,pra, twocolumn,groupaddress,twocolumn,10pt,nofootinbib]{revtex4-1} 
\usepackage{amsmath}
\usepackage{latexsym}
\usepackage{graphicx}
\usepackage{amsfonts}
\usepackage{hyperref}
\usepackage{natbib}
\usepackage{amssymb}
\usepackage[utf8]{inputenc}
\usepackage{dsfont}
\usepackage{subfigure}
\usepackage{paralist}
\usepackage[normalem]{ulem}
\usepackage{braket}
\usepackage{bbold}

\hypersetup{bookmarksopen=true, bookmarksopenlevel=1, unicode=false, pdftoolbar=true, pdfmenubar=true, pdffitwindow=false, pdfstartview={FitH},  pdfauthor={Anette Messinger}, colorlinks=true, linkcolor=blue, citecolor=blue, urlcolor=blue}

\newcommand{\bk}{\mathbf{k}}
\newcommand{\mb}[1]{\mathbf{#1}}
\newcommand{\bs}[1]{\boldsymbol{#1}}
\newcommand{\tm}[1]{\text{#1}}

\newcommand{\curl}[1]{\boldsymbol{\nabla}_{#1}\times}

\newcommand{\xx}{\left(\mathbf{x}-\mathbf{x}'\right)}
\newcommand{\xxw}{\left(\mathbf{x},\mathbf{x}',\omega\right)}
\newcommand{\intk}{\int \frac{d^3k}{(2\pi)^3}\;}

\renewcommand{\Im}{\text{Im}}

\newcommand{\e}{\uuline{\epsilon}}

\newcommand{\M}{\uuline{M}}
\newcommand{\N}{\uuline{N}}
\newcommand{\G}{\uuline{G}}
\newcommand{\id}{\mathbb{1}}

\makeindex

\begin{document}

\title{Spontaneous emission in anisotropic dielectrics}

\author{Anette Messinger}
\email{a.messinger.1@research.gla.ac.uk}
\author{Niclas Westerberg}
\author{Stephen M. Barnett}
\affiliation{School of Physics and Astronomy, University of Glasgow,  Glasgow G12 8QQ, United Kingdom}

\makeatletter
\def\Dated@name{Date: }
\makeatother
\date{\today}

\begin{abstract}
The emission properties of atoms lie at the foundations of both quantum theory and light-matter interactions. In the context of macroscopic media, exact knowledge thereof is important both in current quantum technologies as well as in fundamental studies. While for isotropic media, this is a very well-studied problem, there are still big gaps in the theory of anisotropic media. In particular, to the best of our knowledge, an explicit expression for the spontaneous emission rate in general anisotropic media has not been presented. In this work, we first derive the quantised electromagnetic field operators to calculate the emission rate in uniaxial media. For the more general case of biaxial media we propose an approximate expression based on interpolation between the limiting cases of uniaxial media. We support our model with numerical simulations which are in strong agreement for typical media configurations, and furthermore show how local field effects can be taken into account in the model.
\end{abstract}

\maketitle

\section{Introduction}
The interaction of light with matter has been extensively studied in the past century, both in classical \cite{lorentz16,born33} and quantum mechanical frameworks \cite{loudon00,cohen,walls}. With the recent advance of quantum technologies, the ability to exactly control and predict the behaviour of atoms, or artificial qubits, has become of great importance \cite{yablonovitch,chikkaraddy,lodahl}. Especially the effect of host materials plays a crucial role in numerous solid-state set-ups where one wishes to isolate and control specific impurity atoms or molecules in a medium, examples of which include nitrogen/silicon-vacancy (NV/SV) centres \cite{Kane98,Robledo11,Dutt07}, dye molecules in anthracene \cite{Wang19}, and quantum dots \cite{lodahl}. A special case of such host media which are commonly used are anisotropic crystals. Anisotropic materials cannot be described by a scalar electric permittivity, as they have different responses to the electric field depending on its direction. Apart from crystals where the anisotropy comes naturally from the crystal structure, anisotropic effects also occur in the newly emerging field of metamaterials \cite{engheta06,enghetaBook,solymar09}, where novel macroscopic electromagnetic properties are obtained from a discrete set of artificial elements mimicking the atoms of a medium, a setting which has recently attracted some attention with regards to the spontaneous emission properties \cite{BoydEngheta}. Especially when the effective medium properties are obtained by layering different materials there can be a large anisotropy with respect to the direction of the layers. It is with these kinds of media that we are concerned in this paper, in particular, the modification of the spontaneous dipole emission rate of atoms embedded in anisotropic media.

The rate of spontaneous emission can {readily} be calculated by Fermi's golden rule \cite{fermi50,loudon00}. It has been found that the emission rate is not an intrinsic property of the atom alone, but it also depends on the form of the electromagnetic (vacuum) field it interacts with, which can be modified by the environment \cite{purcell46, drexhage74, haroche06}. For an atom in isotropic bulk media, this modification is given by the refractive index of the medium $n$, such that $\gamma = n\gamma_\tm{vac}$ with $\gamma_\tm{vac}$ being the vacuum decay rate \cite{Barnett92}. In anisotropic media however, the refractive index varies with the electric field direction and therefore the effect on the spontaneous emission rate is more complicated. Not only is the mode density different in such media, but also the propagation of waves themselves, as the wave velocity now also depends on the polarisation and propagation direction \cite{New13,landau13,braat19,born13,new11}, from which phenomena such as birefringence emerge. {This in turn influences the local density of states of the electromagnetic vacuum, and we expect a more complex spontaneous emission rate.} The special case of uniaxial media, which have the same optical properties {in} two orthogonal directions, has already received considerable attention in literature \cite{Chen73,Weiglhofer90,Weiglhofer93,Perinova05,chance78}, as the wave equations are relatively easy to solve due to the additional symmetry.
However, an explicit form of the spontaneous emission rate in such media has not been reported yet\footnote{We note here that Ref.~\cite{chance78} report a result for the radiative lifetime of atoms in uniaxal media in their appendix, however, a quick check of the derivation reveals what seems to be an error in their final result.}. In biaxial media, although considerable effort has gone into characterizing the wave properties \cite{New13, Maldonado91,yariv83,jenkins65,lovett89}, the solutions are complicated enough to make analytical calculations intractable, including an explicit expression for the spontaneous emission rate.

In this work, we quantize the electromagnetic field inside the medium and use this to derive the spontaneous emission rate for an atom in a dielectric medium with arbitrary real permittivity tensor. For uniaxial media we give a closed form of the emission rate for arbitrary dipole alignment. We furthermore propose a model that approximates the emission rate in biaxial media by a linear interpolation between the two limiting cases of uniaxial media (i.e. whether the special anisotropy axis is the same as the dipole direction or orthogonal to it).

Our paper is structured as follows: In {Section}~\ref{sec:quantization} we derive the wave equation in anisotropic dielectrics and quantize the electromagnetic field in terms of solutions of this wave equation. {This is followed by Section}~\ref{sec:Uniaxial-media}, {in which} we calculate the spontaneous emission rate for the easier case of uniaxial media. In Section \ref{sec:Greens}, we discuss an alternative approach using Green's functions. {Additionally, in S}ection~\ref{sec:Biaxial-media} we to propose a model for the rate in general biaxial media based on interpolation between two possible uniaxial limits. We furthermore calculate the rate numerically for certain media configurations and compare the results with those from our model. Finally, in Section~\ref{sec:localfield} we show how local field effects can be taken into account in our results.

\section{Quantization of the electromagnetic field}\label{sec:quantization}
A general anisotropic dielectric medium \cite{landau13,braat19,born13,saleh19,new11} can be described by a permittivity matrix $\e$ that relates the displacement field $\boldsymbol{D}$ to the electric field $\boldsymbol{E}$ as
\begin{equation}
   \boldsymbol{D}=\e \boldsymbol{E}.
\end{equation}
As a consequence, the displacement field is {no longer} parallel to the electric field. In the following{, we} will consider a coordinate system in which $\e$ is diagonal and define
\begin{equation}
 \e=\left( \begin{array}{ccc}
 \epsilon_{\rm x} & 0 & 0 \\ 
 0 & \epsilon_{\rm y} & 0 \\ 
 0 & 0 & \epsilon_{\rm z}
 \end{array} \right),
\end{equation}
{where we also should note that $\epsilon_0$ will throughout this manuscript denote the permittivity of free space.} We aim to find solutions to Maxwell's equations
\begin{align}
\nabla\cdot\boldsymbol{D}&=0\\
\nabla\cdot\boldsymbol{B}&=0\\
\nabla\times\boldsymbol{E}&=-\frac{\partial\boldsymbol{B}}{\partial t}\\
\nabla\times\boldsymbol{H}&=\frac{\partial\boldsymbol{D}}{\partial t},
\end{align}
in such a medium{, where $\boldsymbol{B}$ is the magnetic flux density, related to the magnetic field $\bs{H} = \bs{B}/\mu_0$ with $\mu_0$ being the permeability of free space}. The resulting wave equation for the electric field \cite{Jackson99} is
\begin{align}
	\nabla\times\left(\nabla\times\boldsymbol{E}\right) & =-\mu_{0}\ddot{\boldsymbol{D}}=-\mu_{0}\e\ddot{\boldsymbol{E}.}\label{eq:curlcurlE}
\end{align}
In an anisotropic dielectric, the electric field $ \boldsymbol{E} $ is no longer divergence-free, so we cannot simply replace the left side of Eq.~\eqref{eq:curlcurlE} by a Laplacian to find the Helmholtz equation, as is usually done for isotropic media. However, given that the medium is spatially homogeneous, we can always introduce a decomposition of the electric field into plane waves,
\begin{equation}
\boldsymbol{E}\left(\bs{r},t\right)=\int d^{3}\boldsymbol{k}\,\boldsymbol{E}_{\boldsymbol{k}}{\rm e}^{i(\boldsymbol{k}\cdot\bs{r}-\omega_{\boldsymbol{k}}t)}
\end{equation}
and write Eq.~(\ref{eq:curlcurlE}) as
\begin{align}
	\boldsymbol{k}\times\left(\boldsymbol{k}\times\boldsymbol{E}_{\boldsymbol{k}}\right) & =-\omega_{\boldsymbol{k}}^{2}\mu_{0}\e\boldsymbol{E}_{\boldsymbol{k}}\label{eq:kxkx}\\
	\Leftrightarrow\frac{1}{\mu_{0}}\e^{-1}\left(k^{2}\boldsymbol{E}_{\boldsymbol{k}}-\boldsymbol{k}(\boldsymbol{k}\cdot\boldsymbol{E}_{\boldsymbol{k}})\right) & =\omega_{\boldsymbol{k}}^{2}\boldsymbol{E}_{\boldsymbol{k}}\nonumber 
\end{align}
This is an eigenvalue problem, where $\boldsymbol{E}_{\boldsymbol{k}}$
and $\omega_{\boldsymbol{k}}^{2}$ are eigenvectors and eigenvalues
of the matrix
\begin{align}\label{eq:M}
	M_{ij}=\frac{1}{\mu_{0}\epsilon_{i}}\left(k^{2}\delta_{ij}-k_{i}k_{j}\right),
\end{align}
where we should note that here the double occurrence of an index does not imply use of a summation convention. From the structure of $\uuline{M}$ we can already note a few properties of its solutions:
\begin{enumerate}
	\item There are no more than two non-trivial solutions (with eigenvalues	$\neq0$)
	\item $\omega_{\boldsymbol{k}\lambda}=\omega_{-\boldsymbol{k},\lambda}$, $\boldsymbol{E}_{\boldsymbol{k},\lambda}\Vert\boldsymbol{E}_{-\boldsymbol{k},\lambda}$
	\item $\boldsymbol{k}\cdot(\e\boldsymbol{E}_{\boldsymbol{k},\lambda})=0$\label{eq:keps}
	\item $\boldsymbol{E}_{\boldsymbol{k},\lambda}\cdot(\e\boldsymbol{E}_{\boldsymbol{k},\lambda'})=0$ for $\omega_{\boldsymbol{k}\lambda}\neq\omega_{\boldsymbol{k},\lambda'}$ \label{eq:epseps}
	\item $\frac{1}{\mu_{0}}(\boldsymbol{k}\times\boldsymbol{E}_{\boldsymbol{k},\lambda})\cdot(\boldsymbol{k}\times\boldsymbol{E}_{\boldsymbol{k},\lambda'})=-\omega_{\boldsymbol{k}\lambda}\omega_{\boldsymbol{k},\lambda'}\boldsymbol{E}_{\boldsymbol{k},\lambda}\cdot(\e\boldsymbol{E}_{\boldsymbol{k},\lambda'})$
\end{enumerate}
Detailed proofs of these statements can be found in {Appendix~\ref{appendix:proof}.}

We can interpret these observations the following way: (1.) is simply the fact that there are two polarisations, (2.) {follows from} the reciprocity of the {spatially} homogeneous medium. (3.) tells us that it is $\boldsymbol{D}=\e\boldsymbol{E}$ and not $\boldsymbol{E}$ that is orthogonal to the wave vector{, which is a consequence of} Gauss's law. Similarly, (4.) means that $\boldsymbol{E}_{\boldsymbol{k},\lambda}\bot\boldsymbol{D}_{\boldsymbol{k},\lambda'}$ for different polarisations. This is important for calculating the energy stored in the electric field, which is proportional to $\boldsymbol{E}\cdot\boldsymbol{D}$. Finally, (5.) draws the connection to the magnetic field, i.e. $\boldsymbol{H}_{\boldsymbol{k},\lambda}\cdot\boldsymbol{B}_{\boldsymbol{k},\lambda'}=\boldsymbol{E}_{\boldsymbol{k},\lambda}\cdot\boldsymbol{D}_{\boldsymbol{k},\lambda'}$. In particular, for different polarisations we have $\boldsymbol{H}_{\boldsymbol{k},\lambda}\bot\boldsymbol{B}_{\boldsymbol{k},\lambda'}$,
although in this case we could as well write $\boldsymbol{B}_{\boldsymbol{k},\lambda}\bot\boldsymbol{B}_{\boldsymbol{k},\lambda'}$
or $\boldsymbol{H}_{\boldsymbol{k},\lambda}\bot\boldsymbol{H}_{\boldsymbol{k},\lambda'}$, as here the magnetic flux density and field are related by the scalar permeability of free space $\mu_0$.

With these solutions, let us write the electric and magnetic fields as 
\begin{align}
	\boldsymbol{E}\left(\bs{r},t\right) =& \int d^{3}\boldsymbol{k}\sum_{\lambda}\boldsymbol{e}_{\boldsymbol{k}\lambda}\nonumber\\
	&\left(A_{\boldsymbol{k}\lambda}{\rm e}^{i(\boldsymbol{k\cdot r}-\omega_{\boldsymbol{k}\lambda}t)}+A_{\boldsymbol{k}\lambda}^{*}{\rm e}^{-i(\boldsymbol{k\cdot r}-\omega_{\boldsymbol{k}\lambda}t)}\right)\\
	\boldsymbol{D}\left(\bs{r},t\right) =& \int d^{3}\boldsymbol{k}\sum_{\lambda}\e\boldsymbol{e}_{\boldsymbol{k}\lambda}\nonumber\\
	&\left(A_{\boldsymbol{k}\lambda}{\rm e}^{i(\boldsymbol{k\cdot r}-\omega_{\boldsymbol{k}\lambda}t)}+A_{\boldsymbol{k}\lambda}^{*}{\rm e}^{-i(\boldsymbol{k\cdot r}-\omega_{\boldsymbol{k}\lambda}t)}\right)\\
	\boldsymbol{B}\left(\bs{r},t\right) =& \int d^{3}\boldsymbol{k}\sum_{\lambda}-\frac{1}{\omega_{\boldsymbol{k}\lambda}}\boldsymbol{k}\times\boldsymbol{e}_{\boldsymbol{k}\lambda}\nonumber\\
	&\left(A_{\boldsymbol{k}\lambda}{\rm e}^{i(\boldsymbol{k\cdot r}-\omega_{\boldsymbol{k}\lambda}t)}+A_{\boldsymbol{k}\lambda}^{*}{\rm e}^{-i(\boldsymbol{k\cdot r}-\omega_{\boldsymbol{k}\lambda}t)}\right)\\
	\boldsymbol{H}\left(\bs{r},t\right) =& \int d^{3}\boldsymbol{k}\sum_{\lambda}-\frac{1}{\mu_{0}\omega_{\boldsymbol{k}\lambda}}\boldsymbol{k}\times\boldsymbol{e}_{\boldsymbol{k}\lambda}\nonumber\\
	&\left(A_{\boldsymbol{k}\lambda}{\rm e}^{i(\boldsymbol{k\cdot r}-\omega_{\boldsymbol{k}\lambda}t)}+A_{\boldsymbol{k}\lambda}^{*}{\rm e}^{-i(\boldsymbol{k\cdot r}-\omega_{\boldsymbol{k}\lambda}t)}\right)
\end{align}
with $\boldsymbol{e}_{\boldsymbol{k}\lambda}=\boldsymbol{E}_{\boldsymbol{k},\lambda}/|\boldsymbol{E}_{\boldsymbol{k},\lambda}|$
being the normalized eigenvectors. Using (2.), (4.) and (5.)  we can calculate the energy stored in the field \cite{Jackson99}
\begin{align}\label{eq:FieldEnergy}
	H & =\frac{1}{2}\int d^3r \;\left[ \boldsymbol{E}\left(\bs{r},t\right)\cdot\boldsymbol{D}\left(\bs{r},t\right)+\boldsymbol{H}\left(\bs{r},t\right)\cdot\boldsymbol{B}\left(\bs{r},t\right)\right] \nonumber\\
	  & =(2\pi)^{3}\int d^{3}\boldsymbol{k}\sum_{\lambda}\boldsymbol{e}_{\boldsymbol{k}\lambda}\cdot\e\boldsymbol{e}_{\boldsymbol{k}\lambda}\left(A_{\boldsymbol{k}\lambda}A_{\boldsymbol{k}\lambda}^{*}+A_{\boldsymbol{k}\lambda}^{*}A_{\boldsymbol{k}\lambda}\right).
\end{align}

We quantize the field by introducing the canonical ladder operators $\hat{a}_{\boldsymbol{k}\lambda}$ and
$\hat{a}_{\boldsymbol{k}\lambda}^{\dagger}$ with commutation relations
\begin{equation} \left[\hat{a}_{\boldsymbol{k}\lambda},\hat{a}_{\boldsymbol{k'}\lambda'}^{\dagger}\right]=\delta(\boldsymbol{k}-\boldsymbol{k'})\delta_{\lambda\lambda'} \end{equation}
and make the replacements
\begin{equation}
A_{\boldsymbol{k}\lambda}^{(*)}\rightarrow\sqrt{\frac{\hbar\omega_{\boldsymbol{k}\lambda}}{2(2\pi)^{3}\boldsymbol{e}_{\boldsymbol{k}\lambda}\cdot\e\boldsymbol{e}_{\boldsymbol{k}\lambda}}}\hat{a}_{\boldsymbol{k}\lambda}^{(\dagger)}
\end{equation}
so that we can write the Hamiltonian in diagonal form
\begin{equation}
\hat{H}=\int d^{3}\boldsymbol{k}\sum_{\lambda}\hbar \omega_{\boldsymbol{k}\lambda} \left(\hat{a}_{\boldsymbol{k}\lambda}^\dagger\hat{a}_{\boldsymbol{k}\lambda}+{\frac {1}{2}}\right).
\end{equation}
With this we can rewrite the electric field operator (and analogously all other operators) as
\begin{align}
	\hat{\boldsymbol{E}}\left(\bs{r},t\right)=&\int d^{3}\boldsymbol{k}\sum_{\lambda}\boldsymbol{e}_{\boldsymbol{k}\lambda}\sqrt{\frac{\hbar\omega_{\boldsymbol{k}\lambda}}{2(2\pi)^{3}\boldsymbol{e}_{\boldsymbol{k}\lambda}\cdot\e\boldsymbol{e}_{\boldsymbol{k}\lambda}}}\nonumber\\
	&\left(\hat{a}_{\boldsymbol{k}\lambda}{\rm e}^{i(\bs{k}\cdot\bs{r}-\omega_{\boldsymbol{k}\lambda}t)}+\hat{a}_{\boldsymbol{k}\lambda}^{\dagger}{\rm e}^{-i(\bs{k}\cdot\bs{r}-\omega_{\boldsymbol{k}\lambda}t)}\right).\label{eq:EFieldAniso}
\end{align}
Note that the biggest difference compared to an isotropic medium is the dependency of the frequency on the polarisation and on the direction of $\boldsymbol{k}$. Furthermore, the prefactor (and therefore the commutator of the electric field operator) has an additional dependency on the direction of the polarisation vectors with respect to the crystal axes, $\boldsymbol{e}_{\boldsymbol{k}\lambda}\cdot\e\boldsymbol{e}_{\boldsymbol{k}\lambda}$.

\section{Uniaxial media\label{sec:Uniaxial-media}}
A special {but important} class of anisotropic media are the uniaxial media, where two of the three permittivities are the same. In this case we set $ \epsilon_x=\epsilon_1$ and {$\epsilon_y=\epsilon_2=\epsilon_z$} so that $\e=\mathrm{diag}(\epsilon_1,\epsilon_2,\epsilon_2)$ \cite{New13}. With this additional symmetry it is easy to find solutions to the eigenvalue problem of Eq.~(\ref{eq:kxkx}). The matrix $\uuline{M}$ now has the (un-normalized) eigenvectors\footnote{We omit normalization of eigenvectors throughout this paper as the normalization factor cancels out in all relevant calculations.}
\begin{equation}
\boldsymbol{e}_{\boldsymbol{k}\mathrm{o}}=\left(\begin{array}{c}
0\\
-k_{3}\\
k_{2}
\end{array}\right),\,\,\boldsymbol{e}_{\boldsymbol{k}\mathrm{e}}=\left(\begin{array}{c}
-\epsilon_2(k_{2}^{2}+k_{3}^{2})\\
\epsilon_1k_{1}k_{2}\\
\epsilon_1k_{1}k_{3}
\end{array}\right)\label{eq:EVuni}
\end{equation}
with corresponding angular frequencies
\begin{align}
\omega_{\boldsymbol{k}\mathrm{o}}=\frac{ck}{n_{\rm o}}=\frac{1}{\sqrt{\mu_{0}\epsilon_2}}k\\
\omega_{\boldsymbol{k}\mathrm{e}}=\frac{ck}{n_{\rm e}}=\sqrt{\frac{\boldsymbol{\kappa}\cdot\e\boldsymbol{\kappa}}{\mu_{0}\epsilon_1\epsilon_2}}k,\label{eq:EWuni}
\end{align}
where $\boldsymbol{\kappa}=\boldsymbol{k}/k$, and $n_{\rm o}$ and $n_{\rm e}$ are the ordinary and extraordinary refractive indices, respectively. The first solution corresponds to the ordinary wave. Its polarisation
vector $\boldsymbol{e}_{\boldsymbol{k}\mathrm{o}}$ is still orthogonal to the wavevector and the frequency $\omega_{\mathrm{o}}$ does not depend on the orientation of $\boldsymbol{k}$, just as we would expect in an isotropic medium. It is only the extraordinary wave, $\boldsymbol{e}_{\boldsymbol{k}\mathrm{e}}$, that exhibits the unusual properties that {originate from} the anisotropy \cite{New13}.

Using the electric field representation of Eq.~(\ref{eq:EFieldAniso}), we can now from Fermi's golden rule \cite{loudon00} calculate the spontaneous emission rate of an atomic dipole with transition frequency $\omega_{A}$ and dipole moment $\boldsymbol{d}=(d_1,d_2\cos\phi,d_2\sin\phi)$, which we may assume to be a constant property of the atom, unaffected by the surrounding medium,
\begin{align}\label{eq:fermi}
\gamma & =\frac{2\pi}{\hbar^{2}}\sum_{f}|\bra{f}\hat{\boldsymbol{d}}\cdot\hat{\boldsymbol{E}}\ket{0}|^{2}\delta(\omega_{\boldsymbol{k}\lambda}-\omega_{A})\nonumber\\
& =\frac{1}{8\hbar\pi^{2}}\int d^{3}\boldsymbol{k}\sum_{\lambda}\frac{\omega_{\boldsymbol{k}\lambda}\left|\boldsymbol{d}\cdot\boldsymbol{e}_{\boldsymbol{k}\lambda}\right|^{2}}{\boldsymbol{e}_{\boldsymbol{k}\lambda}\cdot\e\boldsymbol{e}_{\boldsymbol{k}\lambda}}\delta(\omega_{\boldsymbol{k}\lambda}-\omega_{A}).
\end{align}
As there is nothing distinguishing the $y$-axis and $z$-axis, we choose $\phi=0$ for the dipole orientation without loss of generality. Using the linear dispersion relations given in Eq.~\ref{eq:EWuni}, we can make the substitution $k\rightarrow\omega_{\boldsymbol{k}\lambda}$ with $ dk=n_\lambda  d\omega_{\boldsymbol{k}\lambda}/c$ in spherical coordinates
\begin{equation}
\bs{k}=k(\cos\theta, \sin\theta\cos\varphi, \sin\theta\sin\varphi)^{\rm T}.
\end{equation}
After application of the $\delta$-function, this yields
\begin{align}\label{eq:Gamma-angular}
\gamma=\frac{\omega_{A}^{3}}{8\hbar\pi^{2}}\int_0^{2\pi}  d\varphi \int_0^\pi d\theta \sum_{\lambda}\left(\frac{n_\lambda}{c}\right)^{3}\frac{\left|\boldsymbol{d}\cdot\boldsymbol{e}_{\boldsymbol{k}\lambda}\right|^{2}}{\boldsymbol{e}_{\boldsymbol{k}\lambda}\cdot\e\boldsymbol{e}_{\boldsymbol{k}\lambda}}\sin\theta.
\end{align}
\paragraph*{Contributions from ordinary waves:}

{The component $d_1$ of the dipole does not contribute to this emission rate, as ordinary waves have polarisations in the plane with permittivity $\epsilon_2$ only. We can therefore} write the emission rate due to ordinary waves as 
\begin{equation}
\gamma_{\mathrm{o}}=\frac{d_2^{2}\omega_{A}^{3}}{8\hbar\pi^{2}} \int_0^{2\pi}  d\varphi \int_0^\pi d\theta \;\left(\mu_{0}\epsilon_2\right)^{3/2}\frac{\sin^{3}\theta}{\epsilon_2}, 
\end{equation}
with the solution
\begin{equation}
\gamma_{\mathrm{o}}=\frac{d_2^{2}\omega_{A}^{3}\mu_{0}^{3/2}\epsilon^{1/2}_0}{4\pi\hbar}\sqrt{\epsilon_2} \equiv \frac{d_2^{2}\omega_{A}^{3}\mu_{0}^{3/2}}{4\pi\hbar}n_{\rm o}.
\end{equation}
We note a dependency on the ordinary refractive index {$n_{\rm o} = \sqrt{\epsilon_2/\epsilon_0}$} only, which is just what we would expect for ordinary waves {if compared to an isotropic medium.}

\paragraph*{Contributions from extraordinary waves:}

Extraordinary waves{, on the other hand,} can have components both in the plane of $\epsilon_2$
and along the anisotropy axis of $\epsilon_1$, so we cannot omit any parts of the dipole moment for this calculation. However, products of two different components, $e_{\boldsymbol{k}\mathrm{e}}^{(i)}e_{\boldsymbol{k}\mathrm{e}}^{(j)}$ can be omitted due to the structure of the polarisation vector, because they are anti-symmetric in $k_{i}$ and $k_{j}$ and therefore will cancel out after integration. Consequently, we replace the term $\left|\boldsymbol{d}\cdot\boldsymbol{e}_{\boldsymbol{k}\mathrm{e}}\right|^{2}$ in Eq.~(\ref{eq:fermi}) by {$\left(d_1 e_{\boldsymbol{k}\mathrm{e}}^{(x)}\right)^{2}+\left(d_2 e_{\boldsymbol{k}\mathrm{e}}^{(y)}\right)^{2}$} (i.e. omitting all cross-terms). {This yields}
\begin{align}
\gamma_{\mathrm{e}}=&\frac{\omega_{A}^{3}}{2\hbar(2\pi)^{2}}\int_0^{2\pi}  d\varphi \int_0^\pi d\theta \; \frac{(\mu_{0}\epsilon_1\epsilon_2)^{3/2}\sin\theta }{\epsilon_1\epsilon_2(\epsilon_2\sin^{2}\theta+\epsilon_1\cos^{2}\theta)^{5/2}}\nonumber\\
&\quad\quad\quad\quad\quad\quad\quad \times \;\left[d_2^{2}\epsilon_1^{2}\cos^{2}\theta\cos^{2}\varphi+d_1^{2}\epsilon_2^{2}\sin^{2}\theta\right]\nonumber\\
 =&\frac{\omega_{A}^{3}}{3\pi\hbar}\mu_{0}^{3/2}\left(\frac{d_2^{2}\epsilon_1+4d_1^{2}\epsilon_2}{4\sqrt{\epsilon_2}}\right),
\end{align}
{where we should note that this cannot be expressed as a simple function of the extraordinary refractive index 
\begin{align}\label{eq:n_e}
n_{\rm e}=\epsilon_{0}^{-1/2}\left( \cos^2 \theta / \epsilon_2 + \sin^2 \theta / \epsilon_1 \right)^{-1/2},
\end{align}
which is the effective refractive index of light propagating at an angle of $\theta$. This is in stark contrast with both the ordinary wave contribution, and the emission rate in isotropic media. }

\paragraph*{Total emission rate:}

With this, we can write the total emission rate
\begin{align}\label{eq:uniaxialrate}
\gamma & =\gamma_{\mathrm{o}}+\gamma_{{\rm e}}\nonumber\\
& =\frac{\omega_{A}^{3}\mu_{0}^{3/2}}{3\pi\hbar}\left(\frac{\epsilon_1+3\epsilon_2}{4\sqrt{\epsilon_2}}d_2^{2}+\sqrt{\epsilon_2}d_1^{2}\right).
\end{align}
Surprisingly, for a dipole oriented parallel to the $\epsilon_1$-axis, the emission rate is that of an isotropic medium with permittivity $\epsilon_2$. We note that on first glance this is in disagreement with {Ref.}~\cite{chance78}. On further investigation however, we found what appears to be an error in the final steps of the calculation in Ref.~\cite{chance78}. After taking this into account, our results are in fact in agreement.

\begin{figure*}
	\includegraphics[width=1\textwidth]{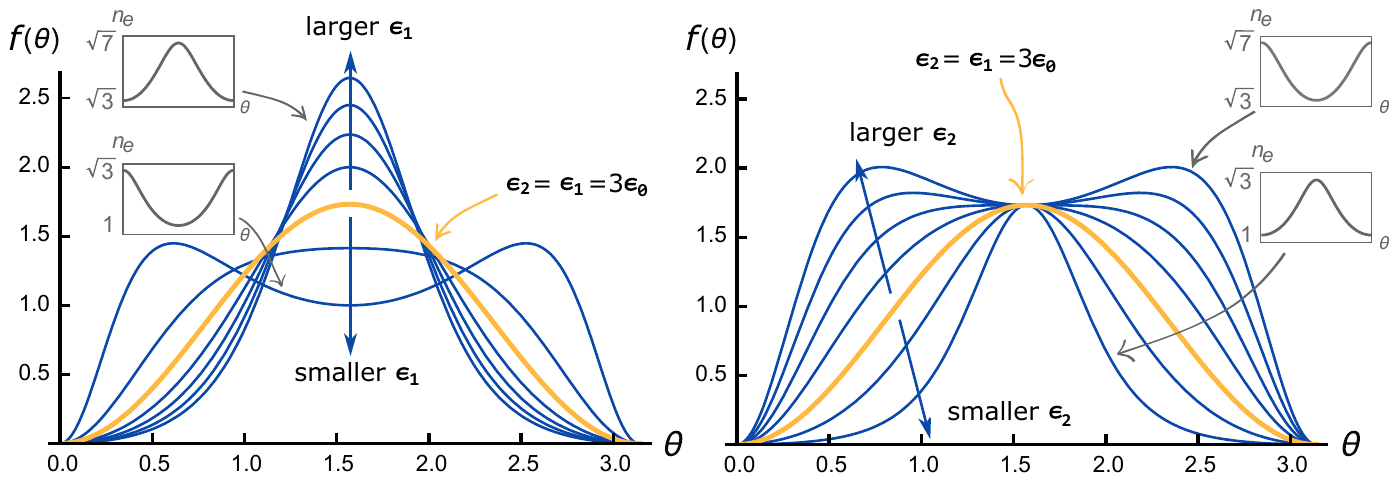}
	\caption{(Color online) Angular distribution $f(\theta)$ of the spontaneous emission rate to a fixed polar angle $\varphi$, for various configurations $\epsilon_{i} \in \{1,7\}$ of a uniaxial medium with fixed $\epsilon_2$ (left) and fixed $\epsilon_1$ (right). A change in $\epsilon_2$ impacts on the amount of radiation to the sides of the distribution, leaving the emission to an angle $\theta=\pi/2$ constant, while a change in $\epsilon_1$ only changes the relative distribution, leaving the total rate (integrated over all angles) constant. The inlays show the extraordinary refractive index of light travelling towards an angle $\theta$ for the two extreme cases of $\epsilon_i=1$ and $7$ respectively.}\label{fig:angular-dependency}
\end{figure*}

To understand our result better, let us look a bit more closely into the emission per unit angle $d\gamma_{\Vert}/d\theta$ by a dipole oriented such that $ \boldsymbol{d}=(d_1,0,0) $. For such a dipole alignment, the emission {couples} purely {to} extraordinary waves. {This is because there} is no dipole component in the plane of $\epsilon_{2}$ and therefore no coupling to the ordinary components of the field. After performing {the first integral} we are left with
\begin{equation}
\gamma_{\Vert}=\frac{\omega_{A}^{3}d_1^{2}}{8\pi^2\hbar}\int_0^{2\pi}  d\varphi \int_0^{\pi}  d\theta \;\frac{\sqrt{\mu_{0}^{3}\epsilon_1\epsilon_2}\sin^{2}\theta\epsilon_2^{2}}{(\epsilon_2\sin^{2}\theta+\epsilon_1\cos^{2}\theta)^{5/2}}\sin\theta
\end{equation}
{from which we can obtain the emission rate per solid angle $d\Omega = d\varphi d\theta \sin\theta$ as}
\begin{align}
\frac{d\gamma_{\Vert}}{d\Omega}&=\frac{\omega_{A}^{3}d_1^{2}\sqrt{\mu_{0}^3}\epsilon_{0}}{8\pi^2\hbar  c^3} \left[\frac{1}{\epsilon_1^2} n_{\rm e}^5(\theta) \sin^{2}\theta\right] \nonumber\\ 
&\equiv {\frac{\omega_{A}^{3}d_1^{2}\sqrt{\mu_{0}^3}\epsilon_{0}}{8\pi^2\hbar  c^3} \left[ f(\theta)/\epsilon_0^2 \right]}
\end{align}
{with $n_{\rm e}$ as given in Eq.~\eqref{eq:n_e}.} We note that the emission to an angle of {$\theta=\pi/2$}, indeed purely depends on $\epsilon_1$, just like we would expect. {It follows that} the dependency of the total rate on $\epsilon_2$ must come due to the effect of the other possible emission directions.  Figure \ref{fig:angular-dependency} shows the angular dependency of the emission rate {$f(\theta)$, }
{with the total emission rate given by}
\begin{align*}
\gamma_{\Vert}~=~\frac{3}{4}\gamma_{{\rm vac}}\int_0^\pi d\theta \; f(\theta)\sin\theta.
\end{align*}
The change in angular distribution can be understood as an interplay between the preferred (orthogonal) dipole emission angle, which arises from the term $\left|\boldsymbol{d}_{\parallel}\cdot\boldsymbol{e}_{\rm \boldsymbol{k}e}\right|^{2}\propto \sin^2\theta$, and the preferred direction of wave propagation towards the minimal optical path length, {which is} determined by the effective refractive index $n_{\rm e}$. Hence, the dipole will predominantly emit towards two azimuthal angles $\theta_{\rm max}=\pi/2 \pm \Delta\theta$ {whenever $\epsilon_2$ is much larger than $\epsilon_1$. In this,} the deviation from orthogonal emission $\Delta\theta=\arccos\sqrt{\frac{2}{3(r-1)}}$ {increases} with the ratio $r=\epsilon_2/\epsilon_1$, while the emission towards {$\theta=\pi/2$}  is fixed by $\epsilon_1$. We note that in fact, the relative angular distribution {$f(\theta)/\left[\int f(\theta)\sin\theta d\theta\right]$} only depends on the ratio $r=\epsilon_2/\epsilon_1$ and not at the product $ \epsilon_2\epsilon_1 $.

\paragraph*{Random dipole orientation}
Finally, we average Eq.~\eqref{eq:uniaxialrate} over random dipole alignments, which leads to the average spontaneous emission rate of unordered emitters
\begin{equation}\label{eq:randomDipole}
	\gamma_{\rm avg} =\frac{\omega_{A}^{3}\mu_{0}^{3/2}d^2}{3\pi\hbar} \left(\frac{1}{6}\frac{\epsilon_{1}}{\sqrt{\epsilon_{2}}}+\frac{5}{6}\sqrt{\epsilon_{2}}\right).
\end{equation}
The lack of an appearance of $\epsilon_{1}$ in the parallel-dipole term is particularly important as it leads to a remarkably weak dependence of the average rate on $\epsilon_{1}$.

\section{Green's function approach\label{sec:Greens}}
Motived by the unexpectedly weak dependence on $\epsilon_1$ in Eq.~\eqref{eq:randomDipole}, it is worth approaching this calculation in an alternative manner. In particular, we are building upon previous work on the dyadic Green's function approach to macroscopic QED \cite{raabe,scheel,Barnett96}, which has successfully been applied to other studies of anisotropic systems in the past \cite{refA0,refA1,refA2,refA3}, and in turn builds on Refs.~\cite{Huttner92,tip,horsley} amongst others. Within this formalism, it can be shown that the decay rate for a dipole at position $\mb{r}_A$ in this formalism is given by 
\begin{align}\label{eq:decayG}
\gamma = {\frac{2\omega_A^2}{\hbar \epsilon_0 c^2}} \mb{d}\cdot\Im\left[\G\left(\mb{r}_A,\mb{r}_A,\omega_A\right)\right]\mb{d}^*,
\end{align}
given that $\G$ is the dyadic Green's function satisfying
\begin{align}\label{eq:Greens}
\curl{}\left(\curl{}\G\xxw \right) - {\omega^2\mu_0}\e\cdot\G\xxw = \id\delta\xx.
\end{align}
As we show in Appendix~\ref{appendix:proofG}, this Green's function can be decomposed in its eigenfunctions as
\begin{align}\label{eq:GreensSol}
\G\xxw = \sum_{\lambda=0}^2 \intk \frac{e^{i\bk\cdot\xx}}{{\mu_0\left(\mb{e}_{\bk\lambda}\cdot\e\cdot\mb{e}_{\bk\lambda}\right)}}\frac{\mb{e}_{\bk\lambda}\otimes\mb{e}_{\bk\lambda}}{\omega^2_{\bk\lambda}-\omega^2},
\end{align}
where we define $\mb{e}_{\bk0} \equiv \bk$ for notational simplicity, along with its eigenvalue $\omega_{\bk 0}=0$. After substituting $\G$ into Eq.~\eqref{eq:decayG}, we arrive at decay rates which are in exact agreement with Eq.~\eqref{eq:Gamma-angular} {(and Eq.~\eqref{eq:biaxialDecay} for biaxial media).}

\section{Biaxial media\label{sec:Biaxial-media}}

\subsection{Wave equation and solutions}
The wave equation of a medium with three different permittivity values,
$\e=\mathrm{diag}(\epsilon_{\rm x},\epsilon_{\rm y},\epsilon_{\rm z})$ has solutions \cite{braat19}
\begin{align}
\boldsymbol{e}_{\boldsymbol{k} \pm} =\left(\begin{array}{c}
k_{1}/(\epsilon_{\rm x}-\epsilon_{\boldsymbol{k} \pm})\\
k_{2}/(\epsilon_{\rm y}-\epsilon_{\boldsymbol{k} \pm})\\
k_{3}/(\epsilon_{\rm z}-\epsilon_{\boldsymbol{k} \pm})
\end{array}\right)\label{eq:evBiaxial}
\end{align}
\begin{align}
\omega_{\boldsymbol{k}\pm}=\frac{ck}{n_\pm}=\frac{1}{\sqrt{\mu_{0}\epsilon_{\boldsymbol{k} \pm}}}k
\end{align}
with 
\begin{align}
\epsilon_{\boldsymbol{k}\pm}=\frac{2\epsilon_{\rm x}\epsilon_{\rm y}\epsilon_{\rm z}}{t_{\boldsymbol{k}}\pm s_{\boldsymbol{k}}},
\end{align}
and $t_{\boldsymbol{k}}=\boldsymbol{\kappa}\cdot\e({\rm Tr}(\e)\mathbf{I}-\e)\boldsymbol{\kappa}$,  $s_{\boldsymbol{k}}=\sqrt{t_{\boldsymbol{k}}^{2}-4\epsilon_{\rm x}\epsilon_{\rm y}\epsilon_{\rm z}\boldsymbol{\kappa}\cdot\e\boldsymbol{\kappa}}$.\footnote{The given representation of the eigenvectors can lead to singularities whenever $\epsilon_{\boldsymbol{k}}$ takes the value of any principal permittivity. This is only a feature of the un-normalized eigenvectors and vanished after normalization.} 
Note that $t_{\boldsymbol{k}}$ and $s_{\boldsymbol{k}}$ depend only
on the direction but not the magnitude of $\boldsymbol{k}$. {The} same is true for the eigenvectors, apart from a constant {pre-factor} $k^{2}$ that will vanish after normalization. With this we can perform the integration over $k$ {in the same manner as before. This yields} the spontaneous emission rate as
\begin{align}\label{eq:biaxialDecay}
\gamma=\frac{\omega_{A}^{3}}{2\hbar(2\pi)^{2}}{\int_0^{2\pi} d\varphi \int_0^\pi d\theta }\sum_{\lambda=\pm}\frac{\left|\boldsymbol{d}\cdot\boldsymbol{e}_{\boldsymbol{\kappa}\lambda}\right|^{2}}{\boldsymbol{e}_{\boldsymbol{\kappa}\lambda}\cdot\e\boldsymbol{e}_{\boldsymbol{\kappa}\lambda}}\left(\mu_{0}\epsilon_{\boldsymbol{k} \pm}\right)^{\frac{3}{2}}\sin\theta.
\end{align}
In the following, we solve the remaining integral numerically, {as well as} introduce a model {that accurately approximates} the solution with an analytical expression based on the known rates in uniaxial media.

\subsection{Dipole along crystal axis -- numerical solution}
\begin{figure}
		\includegraphics[width=\columnwidth]{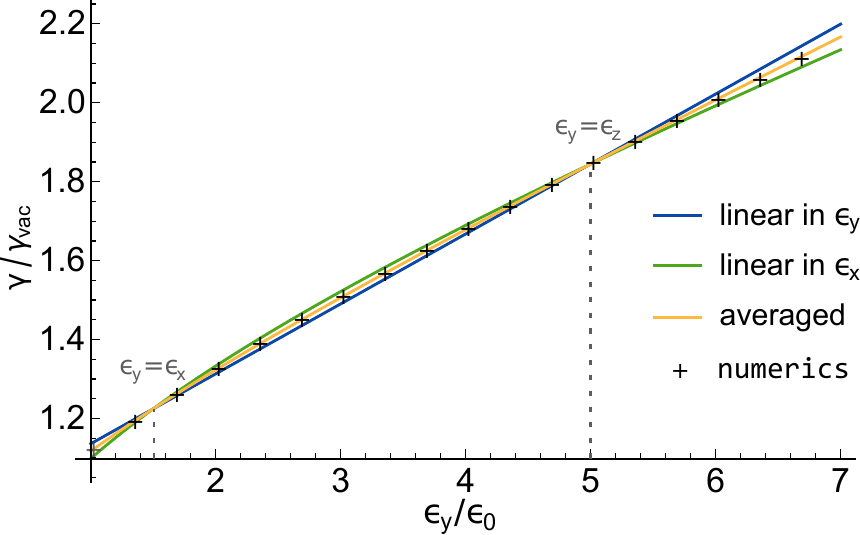}
	\caption{\label{fig:fit}(Color online) Dependency of the spontaneous emission rate (in dimensionless
		units) on the relative permittivity $\epsilon_{\rm y}/\epsilon_{0}$ with
		fixed values of $\epsilon_{\rm x}=1.5\epsilon_{0}$ and $\epsilon_{\rm z}=5\epsilon_{0}$ for a dipole aligned with $\epsilon_{\rm z}$. Analytical models  obtained from linear interpolation with $\epsilon_{\rm x}$, linear interpolation with $\epsilon_{\rm y}$  and an average of both (solid lines) are compared to numerical results (crosses). }
\end{figure}

Let us first consider a dipole aligned in the $z$-direction embedded in the biaxial medium. {Using} the results for uniaxial media, we {note} that if $\epsilon_{\rm y}=\epsilon_{\rm x}$ (i.e. dipole along extraordinary axis), we can identify $\epsilon_{\rm z}$ with the extraordinary index $\epsilon_{\rm 1}$ and $\epsilon_{\rm y}\,\&\,\epsilon_{\rm x}$ with the ordinary index $\epsilon_{\rm 2}$, and the emission rate is given by the $d_1$-part of Eq.~\eqref{eq:uniaxialrate}, {i.e. }
\begin{equation}
\gamma^{(a)}=\frac{d^{2}\omega_{A}^{3}\mu_{0}^{3/2}}{3\pi\hbar}\sqrt{\epsilon_{\rm x}}.
\end{equation}
Similarly, the emission rate is that of a dipole aligned in the $\epsilon_{2}$-plane {if $\epsilon_{\rm y}=\epsilon_{\rm z}$}. {In the same manner,} we can write the rate as
\begin{equation}
\gamma^{(b)}=\frac{d^{2}\omega_{A}^{3}\mu_{0}^{3/2}}{3\pi\hbar}\frac{\left(\epsilon_{\rm x}+3\epsilon_{\rm z}\right)}{4\sqrt{\epsilon_{\rm z}}}.
\end{equation}
If we now fix $\epsilon_{\rm x}$ and $\epsilon_{\rm z}$, we numerically find
nearly linear behaviour with $\epsilon_{\rm y}$ (see Fig.~\ref{fig:fit},
crosses). 
This suggests a linear interpolation between the two known values from the uniaxial cases, 
\begin{align}\label{eq:firstInterp}
	\gamma(\epsilon_{\rm y}) & =\gamma^{(a)}+(\epsilon_{\rm y}-\epsilon_{\rm x})\frac{\gamma^{(b)}-\gamma^{(a)}}{\epsilon_{\rm z}-\epsilon_{\rm x}}\\
	& =\frac{d^{2}\omega_{A}^{3}\mu_{0}^{3/2}}{3\pi\hbar}\left(\sqrt{\epsilon_{\rm x}}-\frac{\epsilon_{\rm y}-\epsilon_{\rm x}}{4\sqrt{\epsilon_{\rm z}}}+\frac{\epsilon_{\rm y}-\epsilon_{\rm x}}{\sqrt{\epsilon_{\rm x}}+\sqrt{\epsilon_{\rm z}}}\right).\nonumber
\end{align}
{This can be seen in} Fig.~\ref{fig:fit}, blue line. However, Eq.~\eqref{eq:firstInterp} is not symmetric with respect to {the exchange of} $\epsilon_{\rm x}$ and $\epsilon_{\rm y}$. As there is nothing distinguishing
$\epsilon_{\rm x}$ and $\epsilon_{\rm y}$ from each other, a similar formula can be written down to be linear in $\epsilon_{\rm x}$ (green line in Fig. \ref{fig:fit}):
\begin{equation}
\gamma(\epsilon_{\rm x})=\frac{d^{2}\omega_{A}^{3}\mu_{0}^{3/2}}{3\pi\hbar}\left(\sqrt{\epsilon_{\rm y}}+\frac{\epsilon_{\rm y}-\epsilon_{\rm x}}{4\sqrt{\epsilon_{\rm z}}}-\frac{\epsilon_{\rm y}-\epsilon_{\rm x}}{\sqrt{\epsilon_{\rm y}}+\sqrt{\epsilon_{\rm z}}}\right)
\end{equation}
Both models deviate from the actual data on two different sides which suggests an average of both. {By t}aking the arithmetic mean, {we find}
\begin{align}\label{eq:average}
	\gamma & =\frac{\gamma(\epsilon_{\rm x})+\gamma(\epsilon_{\rm y})}{2} \\
	& =\frac{d^{2}\omega_{A}^{3}\mu_{0}^{3/2}}{6\pi\hbar}\left[\sqrt{\epsilon_{\rm x}}+\sqrt{\epsilon_{\rm y}}+\frac{\epsilon_{\rm y}-\epsilon_{\rm x}}{\sqrt{\epsilon_{\rm x}}+\sqrt{\epsilon_{\rm z}}}+\frac{\epsilon_{\rm x}-\epsilon_{\rm y}}{\sqrt{\epsilon_{\rm y}}+\sqrt{\epsilon_{\rm z}}}\right].\nonumber
\end{align}
{Thus we arrive at a formula that} is symmetric between $\epsilon_{\rm x}$ and $\epsilon_{\rm y}$ and closely fits the numerical data (see orange curve, Fig.~\ref{fig:fit}).

\begin{figure}
	\includegraphics[width=1\columnwidth]{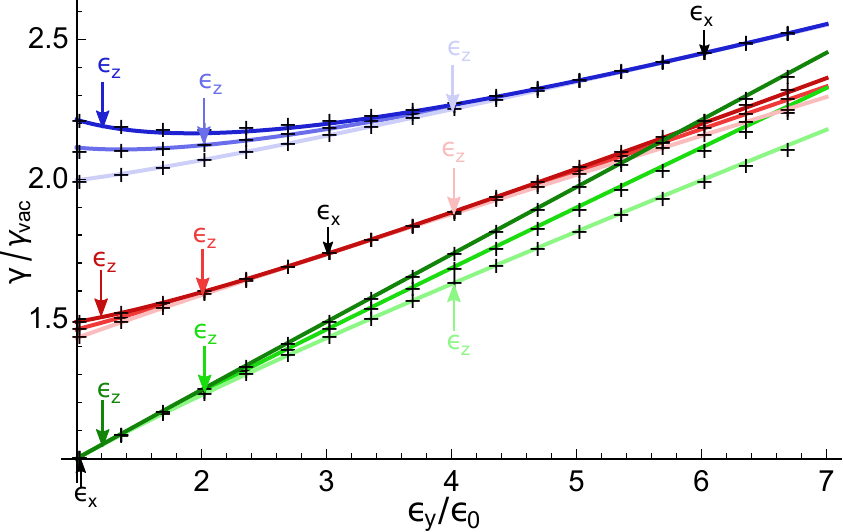}

	\caption{\label{fig:various}(Color online) Comparison of the averaged model (solid lines) with numerical results (crosses) for various different configurations of $\epsilon_{\rm x}/\epsilon_{0}=$ 6, 3, 1 (blue/top, red/middle, green/bottom group of graphs) and $\epsilon_{\rm z}/\epsilon_{0}=$ 4, 2, 1.2 (light, medium, dark graph from each group) for a dipole aligned with $\epsilon_{\rm z}$. The corresponding values for $\epsilon_{\rm x}$ and $\epsilon_{\rm z}$ are also indicated by arrows where they match the value of $\epsilon_{\rm y}$ for each curve.}
	
\end{figure}

{Finally, by i}ntroducing new variables, $n_{+}=\frac{1}{2\sqrt{\epsilon_{0}}}(\sqrt{\epsilon_{\rm y}}+\sqrt{\epsilon_{\rm x}})$,
$n_{-}=\frac{1}{2\sqrt{\epsilon_{0}}}(\sqrt{\epsilon_{\rm y}}-\sqrt{\epsilon_{\rm x}})$
and $n_{\parallel}=\sqrt{\frac{\epsilon_{\rm z}}{\epsilon_{0}}}$ we can
simplify Eq.~(\ref{eq:average}) to
\begin{equation}
\gamma=\left[n_{+}\frac{(n_{+}+n_{\parallel})^{2}+3n_{-}^{2}}{(n_{+}+n_{\parallel})^{2}-n_{-}^{2}}\right]\gamma_{\rm vac} .
\end{equation}
To check the range in which this model is valid, various configurations for $\epsilon_{\rm x}$ and $\epsilon_{\rm z}$ are shown in Fig. \ref{fig:various}. For realistic values, Eq.~(\ref{eq:average}) gives a good approximation to the numerical results. We notice that the permittivity parallel to the dipole axis $\epsilon_{\rm z}$ only weakly influences the emission rate whenever the two orthogonal permittivities $\epsilon_{\rm x}$ and $\epsilon_{\rm y}$ are {of similar size. This is especially the case} compared to its dependency on the other two values.

\subsection{Arbitrary dipole alignment}

If we take a closer look at the form of the electric field vectors in Eq.~(\ref{eq:evBiaxial}), we see that,
just as in the uniaxial case, the product of two different components $ i $ and $ j $ of an eigenvector is always antisymmetric in $k_{i}$ and $k_{j}$. {In other words,} $\boldsymbol{e}_{1}$ does not have any asymmetric parts, {and}
$\boldsymbol{e}_{2}\propto k_{1}k_{2}$ and $\boldsymbol{e}_{3}\propto k_{1}k_{3}$ for both polarisations. Therefore, all cross-terms cancel out in an integration over $\boldsymbol{k}$. {This yields the following expression for a dipole of arbitrary orientation:}
\begin{equation}
	\gamma=\frac{1}{d^{2}}\left(d_{\rm x}^{2}\gamma_{\parallel\epsilon_{\rm x}}+d_{\rm y}^{2}\gamma_{\parallel\epsilon_{\rm y}}+d_{\rm z}^{2}\gamma_{\parallel\epsilon_{\rm z}}\right),\label{eq:dipolealignment}
\end{equation}
where $\gamma_{\parallel\epsilon_{i}}$ is the transmission rate calculated for the dipole aligned to the crystal axis of $\epsilon_{i}$.

\section{Local Field effects}\label{sec:localfield}

We have treated the medium macroscopically so far, assuming that the electric field seen by the dipole is exactly the averaged field over the medium. However, the dipole itself polarizes the surrounding medium, and in order to include such microscopic effects one can introduce a local field correction factor $L$ \cite{inlorentz16,Mossotti50, Onsager36, debye45, vonhippel54} so that $\boldsymbol{E}_{\mathrm{loc}}=L\boldsymbol{E}$, where $\boldsymbol{E}_{\rm loc}$ is the actual field at the dipole position and $\boldsymbol{E}$ is the field according to the macroscopic Maxwell equations.
With this, the spontaneous emission rate requires adjustment. For isotropic media, the corrected spontaneous emission rate is simply
\begin{equation}
\gamma_{\mathrm{loc}}=L^{2}\gamma
\end{equation}
{since $\gamma\propto\left|\boldsymbol{d}\cdot\boldsymbol{E}\right|^{2}$.} In anisotropic media, the correction must also depend on the direction of the electric field, and a reasonable expression would be
\begin{equation}\label{eq:localField}
\boldsymbol{E}_{\mathrm{loc}}=\uuline{L}\boldsymbol{E}
\end{equation}
with the matrix $L_{ij}\equiv L_{i}\delta_{ij}$ diagonal in the basis of the anisotropy axes. The form of local field corrections in anisotropic media is not entirely clear and strongly depends on the model and the configuration of molecules in the medium and the dipole of interest \cite{Rebane03, Lo01, Aubret19, Agranovich74, Aspnes82}. In the following we show how any local field correction can be incorporated into the expressions for the spontaneous emission rate, as long as the effects are linear in the electric field.
For a tensor-valued local field correction, the correction to the spontaneous emission is no longer a simple multiplicative factor, as the $\left|\boldsymbol{d}\cdot\boldsymbol{E}\right|^{2}$ term needs to be replaced by $\left|\boldsymbol{d}\cdot\uuline{L}\boldsymbol{E}\right|^{2}=|\sum d_{i}L_{ij}E_{j}|^{2}$. With this correction, one would have to solve the new integral
\begin{equation}
\gamma_{\mathrm{loc}}=\frac{1}{2\hbar(2\pi)^{2}}\int d^{3}\boldsymbol{k}\sum_{\lambda}\frac{\omega_{\boldsymbol{k}\lambda}\left|\boldsymbol{d}\cdot\uuline{L}\boldsymbol{e}_{\boldsymbol{k}\lambda}\right|^{2}}{\boldsymbol{e}_{\boldsymbol{k}\lambda}\cdot\e\boldsymbol{e}_{\boldsymbol{k}\lambda}}\delta(\omega_{\boldsymbol{k}\lambda}-\omega_{A}).
\end{equation}
However, we can {rewrite this expression and let the matrix $\uuline{L}$} act on the dipole vector to its left, {such that} $\left|(\boldsymbol{d}^{T}\uuline{L})\boldsymbol{E}\right|^{2}\equiv\left|\widetilde{\boldsymbol{d}}\cdot\boldsymbol{E}\right|^{2}$. {This allows us to} substitute the adjusted dipole vector $\widetilde{\boldsymbol{d}}=\uuline{L}^T \boldsymbol{d}$ into the solutions from sections \ref{sec:Uniaxial-media} and \ref{sec:Biaxial-media}. For a local field correction represented by a diagonal matrix $L_{ij}=\delta_{ij}L_i$, we obtain the new expressions for the corrected spontaneous emission rate,
\begin{align}
	\gamma_{\mathrm{loc}} & =\frac{\omega_{A}^{3}\mu_{0}^{3/2}}{3\pi\hbar}\left(\frac{\epsilon_1+3\epsilon_2}{4\sqrt{\epsilon_2}}L_{2}^{2}d_2^{2}+\sqrt{\epsilon_2}L_{1}^{2}d_1^{2}\right)
\end{align}
in a uniaxial medium, and
\begin{equation}
\gamma_{\mathrm{loc}}=\frac{1}{d^{2}}\left(d_{\rm x}^{2}L_{1}^{2}\gamma_{\parallel\epsilon_{\rm x}}+d_{\rm y}^{2}L_{2}^{2}\gamma_{\parallel\epsilon_{\rm y}}+d_{\rm z}^{2}L_{3}^{2}\gamma_{\parallel\epsilon_{\rm z}}\right).
\end{equation}
in a biaxial medium. For each dipole component, the correction is a scalar factor again, so we don't expect any qualitative difference {to present itself, including in the accuracy of the chosen interpolative} model compared to the numerical results.

\section{Conclusion}
In this paper we have quantised the electromagnetic field {in absorption-less} anisotropic dielectrics and used the quantised field operators to derive analytic expressions for the spontaneous emission rate of an electric dipole. {In particular, we found an exact expression in uniaxial media, and we} furthermore presented a simple formula which approximates the spontaneous emission rate in biaxial media. {The latter reduces} to the exact result in the special case of uniaxial media. Our {biaxial} model is in strong agreement with numeric simulations for realistic choices of the principal refractive indices. 

{Interestingly, we found a remarkably weak dependence of the emission rate on the extraordinary permittivity ($\epsilon_1$). Specifically, $\gamma_{\rm avg} \propto \epsilon_2^{-1/2}\left(\epsilon_1/6 + 5\epsilon_2/6\right)$ for a randomly aligned dipole. This should be compared to $\gamma_{\rm avg} \propto n = \sqrt{\epsilon}$ in an isotropic medium. Also, here we note that the spontaneous emission rate cannot be expressed as a simple function of the refractive index in anisotropic media. Both of the above has its roots in the direction-dependence of the extraordinary refractive index $n_{\rm e}(\theta)$, which creates an interplay between the preferred emission direction of the dipole [$\propto \sin^2\theta$] and the favoured direction of propagation of the emitted extraordinary waves [$\propto n_{\rm e}(\theta)$].} 

Additionally, we showed that it is straightforward to generalize the expressions to arbitrary dipole orientations and to include the effects of local field corrections. Due to the simplicity and generality of the model, and at the same time strong agreement to numerical simulations, we expect these results to be of great use for experiments and quantum technology in optical and solid-state set-ups. We wish to highlight that the linear interpolation presented for biaxial media may be of particular use when analysing the specific dependencies of atomic properties on the anisotropic parameters, given that analytical solutions in closed form present a theoretical challenge.

\section{Acknowledgements}
We would like to acknowledge funding from the Engineering and Physical Sciences Research Council under Grant Numbers EP/N509668/1 and EP/R513222/1, as well as The Royal Society under Grant Numbers RP/EA/180010 and RP/150122. We also thank Ed Hinds and Alex Clark for discussions that lead us to this problem.

\appendix
\section{Proofs of electric field properties}\label{appendix:proof}
We present brief proofs of the five properties of the plane-wave electric field solutions given in section \ref{sec:quantization}:
\begin{enumerate}
	\item It can be shown by explicit calculation that ${\rm Rank}(\M)\leq2$.
	\item Equality of forward and backward frequencies follows from the symmetry
	of Eq.~(\ref{eq:kxkx}). The eigenvectors are identical apart from
	an arbitrary prefactor. 
	\item This can be seen from Eq.~(\ref{eq:kxkx}), where the left side
	is clearly orthogonal to $\boldsymbol{k}$.
	\item The matrix $\uuline{M}$ is a product of the diagonal matrix $\e^{-1}$
	and the symmetric matrix $\N:=\frac{1}{\mu_{0}}\left(k^{2}\mathbb{1}-\boldsymbol{k\,k}^{\top}\right)$.
	For a fixed $\boldsymbol{k}$ (we omit that index in the following
	as it is not relevant), we can write
	\begin{align}
	\e^{-1}\N\boldsymbol{E}_{1} & =\omega_{1}^{2}\boldsymbol{E}_{1}\\
	\Leftrightarrow\N\boldsymbol{E}_{1} & =\omega_{1}^{2}\e\boldsymbol{E}_{1}\\
	\Leftrightarrow(\N\boldsymbol{E}_{1})\cdot\boldsymbol{E}_{2} & =\omega_{1}^{2}(\e\boldsymbol{E}_{1})\cdot\boldsymbol{E}_{2}\\
	\Leftrightarrow\boldsymbol{E}_{1}\cdot(\N\boldsymbol{E}_{2}) & =\omega_{1}^{2}\boldsymbol{E}_{1}\cdot(\e\boldsymbol{E}_{2})
	\end{align}
	where in the last line we made use of the fact that both $\N$ and
	$\e$ are symmetric. For the second solution $\boldsymbol{E}_{2}$,
	we know that $\N\boldsymbol{E}_{2}=\omega_{2}^{2}\e\boldsymbol{E}_{2}$
	and therefore, 
	\begin{equation}
	\omega_{2}^{2}\boldsymbol{E}_{1}\cdot(\e\boldsymbol{E}_{2})=\omega_{1}^{2}\boldsymbol{E}_{1}\cdot(\e\boldsymbol{E}_{2}).
	\end{equation}
	So for two different solutions $\omega_{1}\neq\omega_{2}$ we must
	have $\boldsymbol{E}_{1}\cdot(\e\boldsymbol{E}_{2})=0$.
	\item We know that $-\omega_{\boldsymbol{k}\lambda}^{2}\mu_{0}\e\boldsymbol{E}_{\boldsymbol{k},\lambda}=\boldsymbol{k}\times\boldsymbol{k}\times\boldsymbol{E}_{\boldsymbol{k},\lambda}$
	for solutions $\boldsymbol{E}_{\boldsymbol{k},\lambda}$ and $\omega_{\boldsymbol{k}\lambda}$.
	Multiplying a second solution $\boldsymbol{E}_{\boldsymbol{k},\lambda'}$
	from the left, we get
	\begin{align}
	-\omega_{\boldsymbol{k}\lambda}^{2}\mu_{0}\boldsymbol{E}_{\boldsymbol{k},\lambda'}\cdot(\e\boldsymbol{E}_{\boldsymbol{k},\lambda}) & =\boldsymbol{E}_{\boldsymbol{k},\lambda'}\cdot\left(\boldsymbol{k}\times(\boldsymbol{k}\times\boldsymbol{E}_{\boldsymbol{k},\lambda})\right)\\
	& =(\boldsymbol{k}\times\boldsymbol{E}_{\boldsymbol{k},\lambda})\cdot(\boldsymbol{k}\times\boldsymbol{E}_{\boldsymbol{k},\lambda'}).
	\end{align}
	This is nearly what we wanted to show, apart from the prefactor $\omega_{\boldsymbol{k}\lambda}^{2}$.
	For $\lambda=\lambda'$, we have $\omega_{\boldsymbol{k},\lambda'}=\omega_{\boldsymbol{k}\lambda}$
	and we are done. For $\omega_{\boldsymbol{k},\lambda'}=\omega_{\boldsymbol{k}\lambda}$,
	we have shown that $\boldsymbol{E}_{\boldsymbol{k},\lambda}\cdot(\e\boldsymbol{E}_{\boldsymbol{k},\lambda'})=0$,
	so the prefactor does not matter.
\end{enumerate}

\section{Details on Green's function approach}\label{appendix:proofG}

\subsection{Calculating the Green's function}
We can naturally solve Eq.~\eqref{eq:Greens} directly in the form presented. This is the method most commonly employed, see for instance Ref.~\cite{scheel}. However, to better connect this approach to the main body of work, let us first solve the related Green's function $\G'$, as in 
\begin{align}\label{eq:Greensp}
{\frac{1}{\mu_0}}\e^{-1}\left[\curl{}\left(\curl{}\G'\xxw\right)\right] - \omega^2\G'\xxw \nonumber \\
 = \id\delta\xx.
\end{align}
Furthermore, we can solve this using an eigenfunction expansion \cite{barton}, such that
\begin{align}
\G'\xxw = \sum_n \frac{\mb{e}^*_n(\mb{x}')\otimes\mb{e}_n(\mb{x})}{\gamma_n}
\end{align}
where $n$ is the discrete/continuous label for the eigenfunctions $\mb{e}(\mb{x})$ with eigenvalues $\gamma$ satisfying 
\begin{align}\label{eq:eigen}
{\frac{1}{\mu_0}}\e^{-1}\curl{}\left(\curl{}\mb{e}_n(\mb{x})\right) - \omega^2\mb{e}_n(\mb{x}) = \gamma_n\mb{e}_n(\mb{x}).
\end{align}
We can then rewrite Eq.~\eqref{eq:eigen} as
\begin{align}\label{eq:eigen1}
-{\frac{1}{\mu_0}}\e^{-1}\mb{k}\times\left(\mb{k}\times\mb{e}_\mb{k}\right) = \left(\gamma_n+\omega^2\right)\mb{e}_\mb{k},
\end{align}
where we also expanded 
\begin{align}
\mb{e}_n(\mb{x}) = \intk e^{i\mb{k}\cdot\mb{x}} \mb{e}_\mb{k}.
\end{align}
We can now recognise from Eq.~\eqref{eq:M}, i.e. 
\begin{align}
-{\frac{1}{\mu_0}}\e^{-1}\mb{k}\times\mb{k}\times \equiv \M.
\end{align}
Importantly, $\M$ has the eigenvalues and eigenvectors as previously found, i.e.
\begin{align*}
&\mb{k} \text{ with eigenvalue } 0, \\
&\mb{e}_{\bk 1} \text{ with eigenvalue } \omega_{\bk1}, \\
&\mb{e}_{\bk 2} \text{ with eigenvalue } \omega_{\bk2}.
\end{align*}
Here we assume that the eigenvectors are normalised. However, note that we also have to keep track of the null vector $\mb{k}$. Suppose we label $\omega_{\bk0} = 0$ along with $\mathbf{e}_{\bk0} = \boldsymbol{\kappa}$ for notational simplicity, then it is clear that Eq.~\eqref{eq:eigen1} has solutions $\mb{e}_{\bk\lambda}$ for $\lambda = \left\{0,1,2\right\}$ with
\begin{align}
\omega_{\bk\lambda}^2 = \gamma_n+\omega^2\Rightarrow \gamma_{\bk\lambda} = \omega_{\bk\lambda}^2-\omega^2,
\end{align}
where we have identified the index $n$ with the continuous wavevector $\bk$ and the discrete polarisation label $\lambda$. Hence, we find the Green's function for the associated $\G'$-problem as
\begin{align}
\G'\xxw = \sum_{\lambda=0}^2 \intk e^{i\bk\cdot\xx}\frac{\mb{e}_{\bk\lambda}\otimes\mb{e}_{\bk\lambda}}{\omega^2_{\bk\lambda}-\omega^2},
\end{align}
which decomposes in terms of the polarisation vectors $\mb{e}_{\bk1}$ and $\mb{e}_{\bk2}$ along with the wavevector $\bk$. This is however not the Green's function that we need for Eq.~\eqref{eq:decayG}, despite its expedient physical interpretation. To proceed, let's compare Eq.~\eqref{eq:Greens} and Eq.~\eqref{eq:Greensp}. In particular, we want to find $\G$ such that
\begin{align}
\curl{}\left(\curl{}\G\right) = {\frac{1}{\mu_0}}\e^{-1}\left[\curl{}\left(\curl{}\G'\right)\right],
\end{align}
which when written in momentum-space, and multiplied by $\id = \e\;\e^{-1}$, becomes
\begin{align}\label{eq:relateGG}
\e\;\e^{-1}\left[\bk\times\left(\bk\times\uuline{G_\bk}\right)\right] &= {\frac{1}{\mu_0}}\e^{-1}\left[\bk\times\left(\bk\times\uuline{G'_\bk}\right)\right] \\
&\Leftrightarrow {\mu_0}\e \;\M \; \uuline{G_\bk} = \M \;\uuline{G'_\bk}.
\end{align}
From the structure of Eq.~\eqref{eq:relateGG}, we see that Eq.~\eqref{eq:GreensSol} is a viable candidate for $\G$. If we substitute Eq.~\eqref{eq:GreensSol} into Eq.~\eqref{eq:Greens}, it is readily verifiable that
\begin{align}
\curl{}&\left(\curl{}\G\xxw\right) - \omega^2{\mu_0}\e\cdot\G\xxw = \nonumber\\
&{\mu_0}\e \left[\sum_{\lambda=0}^2 \intk e^{i\bk\cdot\xx}\frac{\mb{e}_{\bk\lambda}\otimes\mb{e}_{\bk\lambda}}{{\mu_0}\left(\mb{e}_{\bk\lambda}\cdot\e\cdot\mb{e}_{\bk\lambda}\right)}\right],
\end{align}
which follows from the construction of $\mb{e}_{\bk\lambda}$. Finally, as the polarisation vectors $\mb{e}_{\bk 1}$ and $\mb{e}_{\bk 2}$ along with the wave vector $\bk$ span $\mathbb{R}^3$, such that $\sum_{\lambda=0}^{2} \mb{e}_{\bk\lambda}\otimes\mb{e}_{\bk\lambda} = \id$, it follows that 
\begin{align}
\sum_{\lambda=0}^{2} \frac{\mb{e}_{\bk\lambda}\otimes\mb{e}_{\bk\lambda}}{\mb{e}_{\bk\lambda}\cdot\e\cdot\mb{e}_{\bk\lambda}} = \e^{-1}.
\end{align}
{Here we also used properties~\ref{eq:keps} and \ref{eq:epseps} of the eigenvectors of $M$ already noted in the main manuscript.} We have thus shown that 
\begin{align}
\curl{}&\left(\curl{}\G\xxw\right) - \omega^2{\mu_0}\e\cdot\G\xxw = \nonumber\\ 
&\id \intk e^{i\bk\cdot\xx} = \id\delta\xx,
\end{align}
as intended. Note that for all this, we do need to keep the wavevector $\bk$ in the sum over ``polarisation" vectors. The significance of this is that we technically have an extra soft-photon (longitudinal) decay channel.

\subsection{The decay rate}
Let us ignore the longitudinal response for now, and explore the usual response, so restrict the sum over $\lambda$ to $\{1,2\}$. We then want to find the imaginary part of the Green's function $\G\left(\mb{r}_A,\mb{r}_A,\omega_A\right)$, which is a consequence of the fluctuation-dissipation theorem \cite{Huttner92,scheel}. Using a partial fraction expansion of $\left(\omega^2_{\bk\lambda}-\omega^2\right)^{-1}$ along with the real line version of the Sokhotski-Plemelj theorem \cite{sokhotskii,plemelj1,plemelj2}, we can rewrite
\begin{align}
\Im\bigg[&\frac{1}{\omega^2_{\bk\lambda}-\omega_A^2}\bigg] \\\
&=  \Im\left[\frac{1}{2\omega_A\left(\omega_{\bk\lambda}-\omega_A\right)}-\frac{1}{2\omega_A\left(\omega_{\bk\lambda}+\omega_A\right)}\right] \nonumber\\
&= \frac{1}{2\omega_A}\Im\left[i\pi\delta(\omega_{\bk\lambda}-\omega_A)+\mathcal{P}\frac{1}{\omega_{\bk\lambda}-\omega_A}\right] \nonumber \\
&= \frac{\pi}{2\omega_A}\delta(\omega_{\bk\lambda}-\omega_A),\nonumber
\end{align}
where $\mathcal{P}$ denotes Cauchy's principal value, and we have assumed that $\omega_{\bk\lambda}\geq 0$ (i.e. the second partial fraction does not contribute). We here also assume that all quantities are real except for a small part of $\omega_{\bk\lambda}$ used to choose the right pole. Furthermore, we can rewrite
\begin{align}
\intk &= \int_0^{2\pi}d\varphi\int_{0}^{\pi} d\theta \int_0^\infty dk \; k^2\sin\theta  \\
&= \int_0^{2\pi} d\varphi\int_{0}^{\pi} d\theta \sin\theta \int_{0}^{\infty} \; \frac{d\omega_{\bk\lambda}}{(2\pi)^3} {\left(\frac{n^3_\lambda}{c^3}\right)}\omega^2_{\bk\lambda},\nonumber
\end{align}
where we used that we can in general write 
\begin{align}
\omega_{\bk\lambda} \equiv c k/n_\lambda
\end{align}
With the above in mind, we find that
\begin{align}
&\Im\bigg[\G\left(\mb{r}_A,\mb{r}_A,\omega_A\right)\bigg] \\
&= \Im\left[\sum_{\lambda=1}^2 \intk \frac{e^{i\bk\cdot\left(\mb{r}_A-\mb{r}_A\right)}}{{\mu_0}\left(\mb{e}_{\bk\lambda}\cdot\e\cdot\mb{e}_{\bk\lambda}\right)}\frac{\mb{e}_{\bk\lambda}\otimes\mb{e}_{\bk\lambda}}{\omega^2_{\bk\lambda}-\omega_A^2}\right] \nonumber\\
&= \frac{1}{4\pi^2\mu_0}\sum_{\lambda=1}^2 \int_{0}^{2\pi}d\varphi\int_0^\pi d\theta \sin\theta \; {\left(\frac{n^3_\lambda}{c^3}\right)} \omega_A \frac{\mb{e}_{\bk\lambda}\otimes\mb{e}_{\bk\lambda}}{\mb{e}_{\bk\lambda}\cdot\e\cdot\mb{e}_{\bk\lambda}}.\nonumber
\end{align}
Finally, after substituting this into Eq.~\eqref{eq:decayG}, we find
\begin{align}
\gamma = \frac{\omega_A^3}{8{\hbar} \pi^2} \sum_{\lambda=1}^2 \int_{0}^{2\pi}d\varphi\int_0^\pi d\theta \sin\theta \; {\left(\frac{n^3_\lambda}{c^3}\right)} \frac{\left|\mb{d}\cdot\mb{e}_{\bk\lambda}\right|^2}{\mb{e}_{\bk\lambda}\cdot\e\cdot\mb{e}_{\bk\lambda}},
\end{align}
in agreement with Eqns.~\eqref{eq:Gamma-angular} and Eq.~\eqref{eq:biaxialDecay} for uniaxial and biaxial media, respectively. Note also that here we have assumed that the permittivity is real, but this can be generalised to a complex permittivity, as this does not change the mathematical form of the Green's function $\G$.

\subsection{The longitudinal component}
{For this component,} we need to evaluate
\begin{align}
&\Im\left[\G^\tm{soft}\left(\mb{r}_A,\mb{r}_A,\omega_A\right)\right] \\
&= \Im\left[-\intk \frac{1}{{\mu_0}\omega_A^2}\frac{\bk\otimes\bk}{\bk\cdot\e\cdot\bk}\right] \equiv 0,\nonumber
\end{align}
where we use the fact that $\e$ is a real matrix in the last step. The longitudinal component of the Green's function hence causes an additional decay channel for absorbing media, but is zero for the non-absorbing media of interest here.



\bibliography{biaxial}
\bibliographystyle{apsrev4-1}

\end{document}